\documentclass[conference, 10pt]{IEEEtran}

\ifCLASSINFOpdf
\else
\fi
\usepackage[cmex10]{amsmath}
\usepackage{graphicx}
\usepackage{color}
\usepackage{placeins}
\usepackage{float}
\usepackage{tabularx,colortbl}
\usepackage{fancyhdr}
\newcommand{\SDA}[0]{\ensuremath{\mathtt{SDA}_d}}
\newcommand{\SDAenhzero}[0]{\ensuremath{\mathtt{SDA0}}}
\newcommand{\SDAenhone}[0]{\ensuremath{\mathtt{SDA1}}}
\newcommand{\SDAenhtwo}[0]{\ensuremath{\mathtt{SDA2}}}
\newcommand{\LSDA}[0]{\mathtt{LSDA}}

\newcommand{\nusers}[0]{\ensuremath{N}}
\newcommand{\thre}[0]{\ensuremath{t}}

\newcommand{\send}[2]{\ensuremath{u_{#1}^{#2}}}
\newcommand{\sendplus}[2]{\ensuremath{\tilde{u}_{#1}^{#2}}}
\newcommand{\sendv}[2]{\ensuremath{\mathbf{u}_{#1}^{#2}}}
\newcommand{\sendm}[1]{\ensuremath{\mathbf{U}^{#1}}}
\newcommand{\sendvplus}[2]{\ensuremath{\mathbf{\tilde{u}}_{#1}^{#2}}}
\newcommand{\sendmcomp}[1]{\ensuremath{\mathbf{\mathbf{U}_{#1,\backg{i}}}}}
\newcommand{\recv}[2]{\ensuremath{y_{#1}^{#2}}}
\newcommand{\recvv}[2]{\ensuremath{\mathbf{y}_{#1}^{#2}}}
\newcommand{\recvm}[1]{\ensuremath{\mathbf{Y}^{#1}}}

\newcommand{\prob}[2]{\ensuremath{p_{#1,#2}}}
\newcommand{\sendprof}[1]{\ensuremath{\mathbf{q}_{#1}}}
\newcommand{\recvprof}[1]{\ensuremath{\mathbf{p}_{#1}}}

\newcommand{\recvprofest}[1]{\ensuremath{\hat{\mathbf{p}}_{#1}}}
\newcommand{\sendprofest}[1]{\ensuremath{\hat{\mathbf{q}}_{#1}}}

\newcommand{\probest}[2]{\ensuremath{\hat{p}_{#1,#2}}}
\newcommand{\backg}[1]{\ensuremath{b}}
\newcommand{\sendfreq}[1]{\ensuremath{f_{#1}}}
\newcommand{\sendfreqm}[0]{\ensuremath{\mathbf{F}}}
\newcommand{\autocorr}[0]{\ensuremath{\mathbf{R}_u}}

\newcommand{\binvar}[1]{\ensuremath{s_{j,#1}}}
\newcommand{\binvarm}[0]{\ensuremath{\mathbf{S}_j}}

\newcommand{\uniformi}[1]{\ensuremath{\mu_{#1}}}
\newcommand{\meanuniformi}[0]{\ensuremath{\bar{\mu}}}

\newcommand{\numberfriends}[0]{\ensuremath{n_f}}

\newcommand{\MSEi}[0]{\ensuremath{\mbox{MSE}_i}}
\newcommand{\MSEim}[0]{\ensuremath{\overline{\mbox{MSE}_i}}}

\begin{document}

%
\title{Meet the Family of Statistical Disclosure Attacks}


\author{\IEEEauthorblockN{Simon Oya\IEEEauthorrefmark{1},
Carmela Troncoso\IEEEauthorrefmark{2} and
Fernando P\'erez-Gonz\'alez\IEEEauthorrefmark{1}\IEEEauthorrefmark{2}}
\IEEEauthorblockA{\IEEEauthorrefmark{1}Signal Theory and Communications Dept., University of Vigo}
\IEEEauthorblockA{\IEEEauthorrefmark{2}Gradiant (Galician R\&D Center in Advanced Telecommunications)}}


\maketitle

\thispagestyle{fancy}
\begin{abstract}
Disclosure attacks aim at revealing communication patterns in anonymous communication systems, such as conversation partners or conversation frequency. In this paper, we propose a framework to compare between the members of the statistical disclosure attack family. We compare different variants of the Statistical Disclosure Attack (SDA) in the literature, together with two new methods; as well as show their relation with the Least Squares Disclosure Attack (LSDA).

We empirically explore the performance of the attacks with respect to the different parameters of the system. Our experiments show that i) our proposals considerably improve the state-of-the-art SDA and ii) confirm that LSDA outperforms the SDA family when the adversary has enough observations of the system.

\end{abstract}



%
\IEEEpeerreviewmaketitle

\section{Introduction}
\label{sec:intro}

Mixes constitute the basic building block of high-latency anonymous communication systems \cite{mixes}. They act as a channel that hides the correspondence between incoming and outgoing messages, thus preventing a potential adversary from unveiling users' communication patterns (e.g. friendships, frequency). 

There exist a wide variety of attacks that compromise the anonymity provided by mixes. In this paper, we revisit a particularly efficient family of attacks which is based on the Statistical Disclosure Attack (SDA) \cite{originalSDA} and propose a framework that allows us to easily compare the attacks when performed on threshold mixes. We revisit Mathewson and Dingledine's generalization of the SDA and propose two new variants that outperform previous work. We also illustrate the relation between the SDA and the Least Squares Disclosure Attack (LSDA).

Additionally, we improve the theoretical analysis of the LSDA in \cite{petsLSDA} and extend it to one of the proposed variants of the SDA, which helps us understand the tradeoffs in performance versus complexity when attacking mixes.

The rest of the paper is organized as follows: we start with a brief overview of the current attacks on threshold mixes in Sect.~\ref{sec:relwork}. In Sect.~\ref{sec:sysmodel}, we introduce our system model and notation and then proceed with our revision of statistical disclosure attacks in Sect.~\ref{sec:revision}. We perform a theoretical analysis of the attacks in Sect.~\ref{sec:mse} and validate our results in Sect.~\ref{sec:eval}. Finally, we conclude in Sect.~\ref{sec:conclusions}.

\section{Previous work}
\label{sec:relwork}

The Disclosure Attack \cite{agrawal03} relies on Graph Theory to reveal the exact set of friends of a user (Alice), seeking for mutually disjoint sets of receivers. This attack is known to be NP-complete but there exist other implementations that speed up the search \cite{kesdoganHS}.

Danezis proposed the Statistical Disclosure Attack (SDA) \cite{originalSDA} as a faster alternative to the Disclosure Attack, which is based on the idea that it is possible to statistically isolate Alice's sending behavior after observing a large amount of her message's sets of receivers. The original SDA is limited to a specific scenario and was extended later to a more general user model and more complex mixing algorithms \cite{mathewsonSDA}.

The Least Squares Disclosure Attack (LSDA) \cite{petsLSDA} models profiling as a least squares problem, minimizing the error between the actual number of output messages and a prediction based on the input messages.

In this work, we present an analysis of the family of statistical disclosure attacks \cite{originalSDA,mathewsonSDA} and the LSDA \cite{petsLSDA}, which share the goal of estimating the sending behavior of the users by combining the set of observations in an appropriate way. Other approaches that we leave out of our work are the Two-Sided SDA (TS-SDA) \cite{tsSDA} and the Reversed SDA (RSDA) \cite{RSDA}, which assume that users reply to messages; the Perfect Matching Disclosure Attack (PMDA) and the Normalized Statistical Disclosure Attack (NSDA) \cite{PMDA}, which exploit that the relationship between sent and received messages is one-to-one; and the Bayesian inference-based approach, Vida \cite{Vida}.

\section{System Model and Notation}
\label{sec:sysmodel}

Throughout the text, we will represent vectors using boldface lowercase characters and matrices using boldface capital letters. We will also use $\mathbf{1}_N$ to refer to the column vector whose $N$ elements are equal to 1, and $\mathbf{1}_{N\times M}$ to the all-ones matrix of size $N\times M$. The superscript $T$ will denote the transposing operation.

\paragraph{System Model}

Our system consists of a population of $\nusers$ users, designated by index $i \in \{1, 2, ... \nusers\}$, which communicate using a threshold mix. The system works as follows: every time a user $i$ in our population wants to send a message to another user $j$, she encrypts the message and sends it to the mix. The mix receives and stores the messages until it has gathered $\thre$ of them. Then, it transforms the messages cryptographically to change their appearance and outputs them in a random order; hence hiding the correspondence between incoming and outgoing messages. We call this process a \emph{round} of mixing, and $\thre$ is the \emph{threshold} of the mix.

We denote the number of messages user $i$ sends in round $r$ by $\send{i}{r}$. We define the column vector containing all the messages sent by user $i$ up to round $\rho$ as $\sendv{i}{}=[ \send{i}{1}, \send{i}{2}, ..., \send{i}{\rho}]^T$, and the matrix of all observed inputs to the mix as $\sendm{}=\left( \sendv{1}{}, \sendv{2}{}, ..., \sendv{\nusers}{} \right)$. Likewise, we denote the number of messages user $j$ receives in round $r$ by $\recv{j}{r}$ and define $\recvv{j}{}=[ \recv{j}{1}, \recv{j}{2}, ... , \recv{j}{\rho}]^T$ and $\recvm{}=\left( \recvv{1}{}, \recvv{2}{}, ..., \recvv{\nusers}{}\right)$. Additionally, we define $\sendplus{i}{r}$ as a binary representation of $\send{i}{r}$, denoting whether there is at least one message sent by user $i$ in round $r$ ($\sendplus{i}{r}=1$) or not ($\sendplus{i}{r}=0$).
We also define, $\sendvplus{i}{}=[\sendplus{i}{1}, \sendplus{i}{2}, ... , \sendplus{i}{\rho}]^T$. 

User $i$ sends messages to their recipients according to her \emph{sender profile} and her \emph{sender frequency}. We define the sender profile of user $i$ as $\sendprof{i}=[ \prob{1}{i}, \prob{2}{i}, ..., \prob{\nusers}{i}]^T$, where $\prob{j}{i}$ models the probability that user $i$ sends a message to user $j$. The sender frequency $\sendfreq{i}$ models the probability that a message arriving to the mix comes from user $i$ ($\sendfreq{i}\geq 0$ for $i=1,2, ..., \nusers$ and $\sum_{i=1}^{\nusers} \sendfreq{i}=1$). We also define the vector $\recvprof{j}=[\prob{j}{1}, \prob{j}{2}, ..., \prob{j}{\nusers}]^T$ which shall come in handy later. 
We make no assumptions on the distribution of each sender profile, other than $\prob{j}{i}\geq 0$ for $i,j=1,2,...,\nusers$ and $\sum_{j=1}^{\nusers} \prob{j}{i}=1$ for $i=1,2,...,\nusers$.

We define the \emph{uniformity} of the sender profile of user $i$ as $\uniformi{i}=1-\sum_{j=1}^{\nusers} \prob{j}{i}^2$. The uniformity $\uniformi{i}$ ranges from $0$, when user $i$ always sends messages to the same contact (i.e. $\prob{k}{i}=1$, $\prob{j}{i}=0$ for $k \in \{1, ..., \nusers\}$ and $j\neq k$, $j=1,...,\nusers$), to $\frac{\nusers-1}{\nusers}$, when she sends messages to all the other users equiprobably.

Finally, we define the \emph{background traffic} of a user $i$ as an aggregate of the traffic generated by all users but $i$. This way, vector $\sendv{\backg{i}}{}$ contains the messages sent by all users but $i$, $\sendv{\backg{i}}{}=\sum_{\substack{k=1\\k\neq i}}^{\nusers} \sendv{k}{}=\mathbf{1}_\rho \cdot t - \sendv{i}{}$. The \emph{background profile} is $\sendprof{\backg{i}}=[\prob{1}{\backg{i}}, \prob{2}{\backg{i}}, ... , \prob{\nusers}{\backg{i}}]^T$ where  $\prob{j}{\backg{i}}=\sum_{\substack{k=1\\k\neq i}}^{\nusers}  \frac{\sendfreq{k}}{1-\sendfreq{i}} \cdot \prob{j}{k}$ and the uniformity of this sender profile is denoted by $\uniformi{\backg{i}}$. In all cases, user $i$ will be clear from the context.

\paragraph{Adversary Model}

We consider a global passive adversary that observes the system during $\rho$ rounds. The adversary observes the identity of the users communicating through the mix and knows all the parameters of the system. We also assume that the adversary is not able to link any messages by their content, i.e. the cryptographic transformations do not leak information.

The goal of the adversary is to infer the sending behavior of the users in the system from the observations, i.e. to obtain an estimator $\probest{j}{i}$ of $\prob{j}{i}$ given the input and output observations $\sendm{}$ and $\recvm{}$. 

\section{Revisiting the Family of Disclosure Attacks}
\label{sec:revision}

\subsection{The Original Statistical Disclosure Attack}

Danezis introduced the original Statistical Disclosure Attack ($\SDA$) in \cite{originalSDA}, which provides an estimator of $\prob{j}{i}$ under the assumptions that the user $i$ does not send more than one message each round and the background traffic for that user is uniform, i.e. $\prob{j}{\backg{i}}=\frac{1}{N}$ for $j=1,2,...,\nusers$.

Danezis claims that, by using the Law of Large Numbers, the mean of the observations $\recv{j}{r}$ in the rounds where $i$ has sent at least one message can be written as
\begin{equation} \label{eq:OSDA_y_first}
 \frac{\sendvplus{i}{T} \recvv{j}{}}{\sendvplus{i}{T} \mathbf{1}_{\rho}} \approx  \prob{j}{i} + (t-1) \cdot \prob{j}{\backg{i}}\,,
\end{equation}
and therefore an estimator for $\prob{j}{i}$ is
\begin{equation}
 \probest{j}{i}^{\SDA} = \frac{\sendvplus{i}{T} \recvv{j}{}}{\sendvplus{i}{T} \mathbf{1}_{\rho}}- (t-1) \cdot \probest{j}{\backg{i}}\,
        \mbox{, with } \probest{j}{\backg{i}}=\frac{1}{N}\,.
\end{equation}

In order to compare $\SDA$ with its variants, note that we can write (\ref{eq:OSDA_y_first}) as
\begin{equation} \label{eq:OSDA_y}
 \sendvplus{i}{T} \recvv{j}{} \approx \sendvplus{i}{T} \mathbf{1}_{\rho} \cdot \prob{j}{i} 
                              + \sendvplus{i}{T} \left( \mathbf{1}_{\rho} \cdot t - \mathbf{1}_{\rho} \right) \cdot \prob{j}{\backg{i}}\,.
\end{equation}

\subsection{Generalized Statistical Disclosure Attack}

Mathewson and Dingledine extended Danezis' attack in \cite{mathewsonSDA}, allowing user $i$ to send multiple messages in a round and estimating the background from the observations.

Using this extension, (\ref{eq:OSDA_y}) becomes
\begin{equation} \label{eq:SDAenh0_y}
 \sendvplus{i}{T} \recvv{j}{} \approx \sendvplus{i}{T} \sendv{i}{} \cdot \prob{j}{i} 
                              + \sendvplus{i}{T} \sendv{b}{} \cdot \prob{j}{\backg{i}}\,,
\end{equation}
where we have just replaced the $\mathbf{1}_\rho$s which referred to the number of messages sent by user $i$ in each round in (\ref{eq:OSDA_y}) with the actual number of messages sent by $i$, $\sendv{i}{}$, and $\mathbf{1}_{\rho}\cdot t-\sendv{i}{}=\sendv{b}{}$. 

The background profile is estimated by computing the average number of messages received by $j$ in the rounds where $i$ does not participate and dividing by the total number of messages exiting the mix each round ($\thre$),
\begin{equation} \label{eq:back_est1}
 \probest{j}{\backg{i}}=\frac{1}{t} \cdot \frac{\left(\mathbf{1}_{\rho} - \sendvplus{i}{}\right)^{T} \recvv{j}{} }{\left(\mathbf{1}_{\rho} - \sendvplus{i}{}\right)^{T} \mathbf{1}_{\rho}}\,.
\end{equation}

We denote this attack by $\SDAenhzero$, whose estimator is
\begin{equation} \label{SDAenh0}
 \probest{j}{i}^{\SDAenhzero} = \frac{\sendvplus{i}{T} \recvv{j}{}}{\sendvplus{i}{T} \sendv{i}{}}
                  - \frac{\sendvplus{i}{T}\sendv{b}{}}{\sendvplus{i}{T} \sendv{i}{}} \cdot \probest{j}{\backg{i}}\,.
\end{equation}

\subsection{Improvements in the Generalized SDA}

The attack described in the previous section performs an average of the outputs in those rounds where user $i$ sends at least one message in order to compute $\probest{j}{i}^{\SDAenhzero}$, giving the same value to those outputs regardless of the actual participation of user $i$. We propose a new estimator, which we denote $\SDAenhone$, that counts the outputs once for every message sent by user $i$, therefore giving more weight to those rounds where the number of messages sent by $i$ is larger.


Using this approach, (\ref{eq:SDAenh0_y}) becomes
\begin{equation} \label{eq:SDAenh1_y}
 \sendv{i}{T} \recvv{j}{} \approx \sendv{i}{T} \sendv{i}{} \cdot \prob{j}{i} 
                              + \sendv{i}{T} \sendv{b}{} \cdot \prob{j}{\backg{i}}\,,
\end{equation}
where we have replaced the vector we used to select the rounds we were taking into account, $\sendvplus{i}{}$, by the vector with the actual number of messages sent by $i$ in each round, $\sendv{i}{}$.

From (\ref{eq:SDAenh1_y}), we get the following estimator,
\begin{equation} \label{SDAenh1}
 \probest{j}{i}^{\SDAenhone} = \frac{\sendv{i}{T} \recvv{j}{}}{\sendv{i}{T} \sendv{i}{}}
                  - \frac{\sendv{i}{T}\sendv{b}{}}{\sendv{i}{T} \sendv{i}{}} \cdot \probest{j}{\backg{i}}\,,
\end{equation}
where $\probest{j}{\backg{i}}$ is estimated as in (\ref{eq:back_est1}).

Note that the idea behind this estimator appears in \cite{mathewsonSDA} applied to other mixing algorithms. The analysis of SDA in \cite{mathewsonSDA} also features the idea of exploiting observations from rounds where user $i$ appears as a sender in order to compute $\probest{j}{\backg{i}}$.


The latter idea inspires our second variant, denoted $\SDAenhtwo$, which uses the observations from \emph{all rounds} to get the background estimation. Following (\ref{eq:SDAenh1_y}), we can write
\begin{equation} \label{eq:SDAenh2_y}
\begin{cases} 
 \sendv{i}{T} \recvv{j}{} = \sendv{i}{T} \sendv{i}{} \cdot \probest{j}{i} 
                              + \sendv{i}{T} \sendv{\backg{i}}{} \cdot \probest{j}{\backg{i}}\\
 \sendv{\backg{i}}{T} \recvv{j}{} = \sendv{\backg{i}}{T} \sendv{i}{} \cdot \probest{j}{i} 
                              + \sendv{\backg{i}}{T}  \sendv{\backg{i}}{} \cdot \probest{j}{\backg{i}}\,.
\end{cases}
\end{equation}

If we define the $\rho \times 2$ matrix $\sendmcomp{i}=\left( \sendv{i}{},  \sendv{\backg{i}}{} \right)$, the new estimator $\probest{j}{i}^{\SDAenhtwo}$ can be obtained by solving
\begin{equation} \label{eq:SDAenh2}
 \left( \begin{array}{c}
 \probest{j}{i}^{\SDAenhtwo} \\
 \probest{j}{\backg{i}}  \end{array} \right) =
 \left( \sendmcomp{i}^T  \sendmcomp{i} \right)^{-1}  \sendmcomp{i}^T  \recvv{j}{}\,.
\end{equation}

\subsection{The Least Squares Disclosure Attack}

The estimator in (\ref{eq:SDAenh2}) uses the information from all outputs when estimating both $\prob{j}{i}$ and $\prob{j}{\backg{i}}$. However, users' profiles are solved independently, compressing information in matrices $\sendmcomp{i}$. We can extend the idea in (\ref{eq:SDAenh2_y}) considering that, when computing the sender profile of $i$, the background is formed by all the users but $i$. In that case, we would have $\nusers$ equations with $\nusers$ unknowns, which are
\begin{equation}
 \sendv{i}{T} \recvv{j}{} = \sendv{i}{T} \sum_{k=1}^{\nusers} \left(\sendv{k}{} \cdot \probest{j}{k}\right)\,,\,\mbox{for}\, i=1,...,\nusers\,.
\end{equation}

Presenting this system in matricial form, we have
\begin{equation}
 \sendm{T} \recvv{j}{} = \sendm{T} \sendm{} \recvprofest{j}\,.
\end{equation}

Therefore, if $\sendm{T} \sendm{}$ is not singular, we obtain the Least Squares Disclosure Attack ($\LSDA$) estimator in \cite{petsLSDA}, 
\begin{equation} \label{eq:LSDA}
  \recvprofest{j}^{\LSDA} = \left(\sendm{T} \sendm{}\right)^{-1} \sendm{T} \recvv{j}{}\,.
\end{equation}

\section{Performance Analysis}
\label{sec:mse}

In this section, we aim at deriving a theoretical expression for the \emph{Mean Squared Error of sender profile $i$}, which we define as $\MSEi\doteq||\sendprof{i}-\sendprofest{i}||^2=\sum_{j=1}^{\nusers} \left(\prob{j}{i}-\probest{j}{i}\right)^2$, for the described estimators. Due to space limitations, we reduce our analysis to  $\SDAenhtwo$ and $\LSDA$. 

We start by deriving an expression of $\mbox{MSE}_i$ in $\LSDA$. In order to do so, we first show, by using the law of total expectation together with $\mbox{E}\left\{ \recvv{j}{}| \sendm{} \right\}=\sendm{} \cdot \recvprof{j}$, that this estimator is unbiased, since
\begin{equation}
 \mbox{E}\{ \recvprofest{j} \}=\mbox{E}\left\{ \mbox{E}\left\{ \recvprofest{j} | \sendm{} \right\} \right\}= \mbox{E}\left\{ \left( \sendm{T} \sendm{} \right)^{-1} \sendm{T} \mbox{E}\left\{ \recvv{j}{} | \sendm{}\right\} \right\}=\recvprof{j}
\end{equation}

Using this fact, along with the law of total variance, we can write the covariance matrix of $\recvprof{j}$ as
\begin{equation} \label{eq:sigmaycond}
 \mathbf{\Sigma}_{\recvprof{j}}=\mbox{E}\left\{ \mathbf{\Sigma}_{\recvprof{j} | \sendm{} } \right\}=\mbox{E}\left\{ \left(\sendm{T} \sendm{} \right)^{-1} \sendm{T} \mathbf{\Sigma}_{\recvv{j}{}|\sendm{}}\sendm{}\left(\sendm{T}\sendm{}\right)^{-1} \right\}
\end{equation}

We model $\{\send{1}{r}, ..., \send{\nusers}{r}\}$ together as a multinomial distribution with $t$ trials and probabilities $\{\sendfreq{1}, ..., \sendfreq{\nusers}\}$. In order to compute (\ref{eq:sigmaycond}), we first assume that the number of observations is large enough, so that we can approximate $\sendm{T}\sendm{}\approx\mbox{E}\{\sendm{T}\sendm{}\}=\autocorr \cdot \rho$, where $\autocorr$ is the autocorrelation matrix of the input process,
\begin{equation}
 \autocorr=\thre\left[\sendfreqm + \left(\thre-1\right)\sendfreqm \mathbf{1}_{\nusers\times \nusers} \sendfreqm\right]
\end{equation}
where $\sendfreqm=\mbox{diag}\{ \sendfreq{1}, ..., \sendfreq{\nusers}\}$.

Applying the matrix inversion lemma, we can write the inverse of this autocorrelation matrix as
\begin{equation} \label{eq:invrx}
 \autocorr^{-1}=\frac{1}{t}\left[\sendfreqm^{-1}-\left(1-\frac{1}{\thre}\right)\mathbf{1}_{\nusers\times \nusers} \right]\,.
\end{equation}

Now that, using $\sendm{T}\sendm{} \approx \autocorr \cdot \rho$, the only term remaining inside $\mbox{E}\{\cdot\}$ in (\ref{eq:sigmaycond}) is $\mbox{E}\{ \sendm{T}\mathbf{\Sigma}_{\recvv{j}{}|\sendm{}} \sendm{}\}$. We model $\recv{j}{r}|\sendm{}$ as the sum of $\nusers$ binomial processes with $\send{i}{r}$ trials and probabilities $\prob{j}{i}$, for $i=1, 2, ..., \nusers$. Let $\binvar{k}=\prob{j}{k}\cdot(1-\prob{j}{k})$ and $\binvarm=\mbox{diag}\{\binvar{1}, ..., \binvar{\nusers}\}$. Then, $\mathbf{\Sigma}_{\recvv{j}{}|\sendm{}}$ is a diagonal matrix whose $(r,r)$-th element is $\left(\mathbf{\Sigma}_{\recvv{j}{}|\sendm{}}\right)_{r,r}=\sum_{k=1}^{\nusers} \send{k}{r} \binvar{k}$. Operating,
\begin{equation} \label{eq:midterm}
\begin{array}{l}
 \mbox{E}\{ \sendm{T}\mathbf{\Sigma}_{\recvv{j}{}|\sendm{}} \sendm{}\}=\\
 \rho \left\{ \sendfreqm \left( \eta_j \thre^{(3)} \mathbf{1}_{\nusers\times\nusers} + \binvarm \mathbf{1}_{\nusers\times\nusers} \thre^{(2)} 
 + \mathbf{1}_{\nusers\times\nusers} \binvarm \thre^{(2)} \right) \sendfreqm \right\}\\
 + \ \rho \left\{\left(\eta_j \thre^{(2)} \mathbf{I}_{\nusers\times\nusers} + \thre \binvarm \right) \sendfreqm \right\}
 \end{array}
\end{equation}
where $\eta_j=\sum_{k=1}^{\nusers} \sendfreq{k} \binvar{k}$ and $\thre^{(n)}=\thre\cdot(\thre-1)\cdot ... \cdot(\thre-n+1)$.

Plugging (\ref{eq:invrx}) and (\ref{eq:midterm}) into (\ref{eq:sigmaycond}) we get an approximation of $\mathbf{\Sigma}_{\recvprof{j}}$. Now, taking each of the diagonal elements of this matrix, which are $\mbox{Var}\{\probest{j}{i}\}$ for $i=1, ..., \nusers$ and adding them along $j$ to obtain $\MSEi=\sum_{j=1}^{\nusers} \mbox{Var}\{ \probest{j}{i} \}$, we finally get
\begin{equation} \label{eq:MSE_lsda}
 \MSEi^{\LSDA}\approx \frac{1}{\rho} \left\{ \left(\sendfreq{i}^{-1}-1\right)\left(1-\frac{1}{\thre}\right)\meanuniformi_{\LSDA} + \frac{\sendfreq{i}^{-1}}{\thre} \cdot \uniformi{i}\right\}
\end{equation}
where $\meanuniformi_{\LSDA}=\sum_{k=1}^{\nusers} \sendfreq{k} \uniformi{k}$ is the average uniformity of the sender profiles.

Following a similar approach, it can be shown for $\SDAenhtwo$ that, when the number of observed rounds is large enough,
\begin{equation} \label{eq:MSE_sda2}
 \MSEi^{\SDAenhtwo}\approx \frac{1}{\rho}  \left\{ \left(\sendfreq{i}^{-1}-1\right)\left(1-\frac{1}{\thre}\right)\meanuniformi_{\SDAenhtwo} + \frac{\sendfreq{i}^{-1}}{\thre} \cdot \uniformi{i}\right\}
\end{equation}
where $\meanuniformi_{\SDAenhtwo}=\sendfreq{i}\uniformi{i} + (1-\sendfreq{i})\uniformi{\backg{i}}$ is the average uniformity considering that there are only two users in the system: the user $i$ and her background. 

Note that the only approximations made to derive (\ref{eq:MSE_lsda}) and (\ref{eq:MSE_sda2}) were $\sendm{T}\sendm{}\approx\mbox{E}\{\sendm{T}\sendm{}\}=\autocorr \cdot \rho$ and its equivalent with matrix $\sendmcomp{i}$. Therefore, these MSE estimators are more accurate as the number of observed rounds is large.

Given the definition of the background sending profile in Sect.~\ref{sec:sysmodel}, it is easy to see that $\meanuniformi_{\SDAenhtwo}\geq \meanuniformi_{\LSDA}$, and therefore $\MSEi^{\SDAenhtwo}\geq \MSEi^{\LSDA}$, where the equality holds only when all users have the same sending profile. This proves that $\LSDA$ will eventually outperform $\SDAenhtwo$ in terms of MSE when the attacker observes the system indefinitely.

\section{Evaluation}
\label{sec:eval}

We evaluate the performance of the attacks in Sect.~\ref{sec:revision} in terms of $\MSEi$, simulating a threshold mix system as described in Sect.~\ref{sec:sysmodel}.\footnote{The simulator, written in Matlab, will be available upon request.}
We exclude $\SDA$ from this evaluation and use its generalization $\SDAenhzero$ instead.

We vary the number of users in the population $\nusers$, the threshold $\thre$, the sending frequencies $\sendfreq{i}$, the number of rounds observed by the attacker $\rho$ and the uniformity of the sending profiles $\uniformi{i}$. 

\subsection{Performance with respect to the uniformity $\uniformi{i}$}

As we have shown in Sect.~\ref{sec:mse}, the uniformity of the sender profiles is a key parameter to show the difference in performance between $\SDAenhtwo$ and $\LSDA$. For simplicity, we assume that each user $i$ has $\numberfriends$ friends to whom she sends messages uniformly, which are users $\mbox{mod}\left( i+k, \nusers\right)$ for $k=0, ..., \numberfriends-1$. This allows us to vary the uniformity of the sender profile of each user with a single parameter: $\uniformi{i}=\frac{1-\numberfriends}{\numberfriends}$. We choose the number of friends $\numberfriends$ from $\{10, 25, 50, 100\}$ and, for each value, perform $100$ repetitions of the experiment.

\begin{figure}[!t]
\centering
\includegraphics[width=0.45\textwidth]{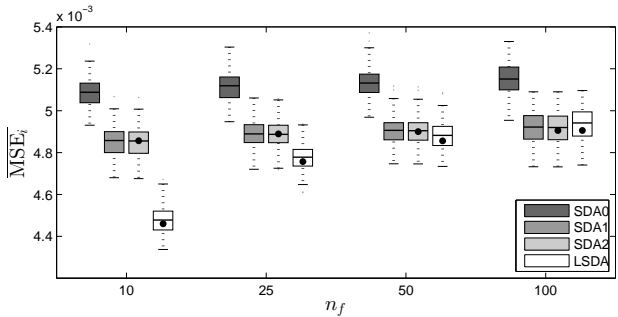}
\caption{Average $\mbox{MSE}$ for the different attacks, as a function of the number of friends $\numberfriends$ of each user ($\rho=20000$, $\nusers=100$, $\sendfreq{i}=1/\nusers$, $t=10$).}
\label{fig:boxplot}
\end{figure}

Figure~\ref{fig:boxplot} shows a box-and-whiskers plot of the average MSE per sender profile, $\MSEim$. On the boxes, the central mark is the mean and the edges are the 25th and 75th percentiles. The black circles $\bullet$ represent the theoretical asymptotic values of the $\MSEim$, from (\ref{eq:MSE_lsda}) and (\ref{eq:MSE_sda2}). Since $\rho$ is finite, $\MSEim$ does not coincide exactly with its theoretical value, although (\ref{eq:MSE_lsda}) and (\ref{eq:MSE_sda2}) reliably describe the accuracy of the attacks. As expected, when the uniformity of the sender profiles is low and the background uniformity $\uniformi{\backg{}}$ is large, $\LSDA$ outperforms the other estimators, but as the uniformity of each user increases and therefore becomes closer to the background uniformity, the advantage of $\LSDA$ decreases. 
Also, note that the proposed estimators $\SDAenhone$ and $\SDAenhtwo$ outperform $\SDAenhzero$.

\subsection{Performance with respect to the other parameters}

Due to space limitations, we are not able to plot the results obtained when varying all the other parameters. We summarize the basic results next and refer to \cite{petsLSDA} for further information about LSDA. First, the $\MSEi$ decreases with $1/\rho$ in each of these attacks, as in (\ref{eq:MSE_lsda}) and (\ref{eq:MSE_sda2}). Also, in every attack, the $\MSEi$ is approximately proportional to the inverse of the sending frequency $\sendfreq{i}^{-1}$, due to the increasing difficulty of estimating the sender profile of a user when she rarely participates in the system. The threshold $\thre$ has little influence on the $\MSEi$ of $\SDAenhtwo$ and $\LSDA$ but does, however, decrease the number of rounds that can be used to estimate the background (\ref{eq:back_est1}) in $\SDAenhzero$ and $\SDAenhone$, thus increasing the $\MSEi$ in these estimators. Finally, we note that increasing $\nusers$ adds an extra error in $\LSDA$ which is not predicted by (\ref{eq:MSE_lsda}) and that stems from the matrix inversion 
in (\ref{eq:LSDA}). This error can be reduced by increasing the number of rounds observed. This can be seen in Fig.~\ref{fig:boxplot}, where the mean values of $\MSEim$ obtained for $\LSDA$ are slightly above their asymptotic value.

The improvements in performance achieved by the more sophisticated versions of statistical disclosure come at the price of an increase in the computational cost. While $\SDAenhzero$ adds the observations where the user whose profile is being estimated has participated, $\SDAenhone$ needs to perform an additional multiplication for each of these rounds. $\SDAenhtwo$ has a higher computational cost since it requires solving a system of two equations for each user, and $\LSDA$ requires solving a linear system of $\nusers$ equations with $\nusers$ unknowns.

\section{Conclusions}
\label{sec:conclusions}

In this work, we have introduced a framework to model the different attacks of the statistical disclosure family, showing how better results can be achieved when performing more complex operations with the observations from the system. We have formalized two new variants of the SDA, which we called $\SDAenhone$ and $\SDAenhtwo$, and showed that they significantly improve the state-of-the-art SDA in threshold mixes, $\SDAenhzero$. Furthermore, we have shown that the LSDA, introduced in \cite{petsLSDA}, can be seen as an upgraded version of statistical disclosure that solves the problem jointly for all users.

We have also improved the previous theoretical analysis on LSDA and derived for the first time an expression which accurately approximates the error of $\SDAenhtwo$. Our experiments confirm these theoretical results.


\section*{\small ACKNOWLEDGEMENTS}
\small
This research was supported by the European Union under
project LIFTGATE (Grant Agreement Number 285901), the European
Regional Development Fund (ERDF) and the Spanish Government
under projects DYNACS (TEC2010-21245-C02-02/TCM) and
COMONSENS (CONSOLIDER-INGENIO 2010 CSD2008-00010),
and the Galician Regional Government under projects ”Consolidation
of Research Units” 2009/62, 2010/85.

\bibliographystyle{IEEEtran}
\bibliography{IEEEabrv,references}
%

\end{document}